\documentclass[notitlepage, 11pt, a4paper]{article}

\usepackage{graphicx}
\usepackage[caption=false]{subfig}
\usepackage{units}

\usepackage{authblk}
\usepackage{amsmath}
\usepackage{textcomp}

\newcommand\blfootnote[1]{%
  \begingroup
  \renewcommand\thefootnote{}\footnote{#1}%
  \addtocounter{footnote}{-1}%
  \endgroup
}

\begin{document}

%emergencystretch: 10pt

\title{Grain-size dependent demagnetizing factors in permanent magnets} 

\author{S. Bance}
\author{B. Seebacher}

\affil{Department of Technology, St P\"{o}lten University of Applied Sciences, Matthias Corvinus-Str. 15, 3100 St P\"{o}lten, Austria}

\author{T. Schrefl}
\affil{Center for Integrated Sensor
Systems, Danube University Krems, 2700 Wiener Neustadt, Austria}
 
\author{L. Exl}
\affil{Solid State Physics,
Vienna University of Technology, 1040 Vienna, Austria}

\author{M. Winklhofer}
\affil{Department of Earth and Environmental Science, Ludwig-Maximilians-Universit\"at M\"unchen, 80333 Munich, Germany}
\affil{Fakult\"at f\"ur Physik, Universit\"at Duisburg-Essen, 47048 Duisburg, Germany}

\author{G. Hrkac}
\affil{CEMPS, Harrison Building, University of Exeter, Exeter, EX4 4QF, UK}

\author{G. Zimanyi}
\affil{Department of Physics and Astronomy, UC Davis, One Shields Avenue, Davis, California 95616, USA}

\author{T. Shoji}
\author{M. Yano}
\author{N. Sakuma}
\author{M. Ito}
\author{A. Kato}
\author{A. Manabe}
\affil{Toyota Motor Corp., Toyota City, 471-8572, Japan}

%\date{\today}
\date{3 October 2014}

\maketitle

\blfootnote{
The following article appeared in S. Bance et al., ``Grain-Size Dependent Demagnetizing Factors in Permanent Magnets.`` J. App. Phys. vol. 116 no. 23 pp. 233903 and may be found at http://scitation.aip.org/content/aip/journal/jap/116/23/10.1063/1.4904854 \\

\textcopyright 2014 American Institute of Physics. This article may be downloaded for personal use only. Any other use requires prior permission of the author and the American Institute of Physics. \\

email:  s.g.bance@gmail.com

web:  http://academic.bancey.com

}

\clearpage

\begin{abstract}
The coercive field of permanent magnets decreases with increasing grain size. The grain size dependence of coercivity is explained by a size dependent demagnetizing factor. In Dy free Nd$_2$Fe$_{14}$B magnets the size dependent demagnetizing factor ranges from 0.2 for a grain size of 55 nm to 1.22 for a grain size of  8300 nm. The comparison of experimental data with micromagnetic simulations suggests that the grain size dependence of the coercive field in hard magnets is due to the non-uniform magnetostatic field in polyhedral grains.

\end{abstract}

%\pacs{75.60.Ch, 96.12.Hg, 75.60.Ej }% insert suggested PACS numbers in braces on next line
%\doi{10.1063/1.4904854}

\section{Introduction} 
Renewable energy technologies heavily rely on permanent magnets.\cite{Gutfleisch2011} Permanent magnets are used in direct drive wind power generators and in the motors and generators of hybrid vehicles. Modern high performance permanent magnets are based on Nd$_2$Fe$_{14}$B. In many applications the magnets are used at temperatures well above room temperature. For example the operating temperature in hybrid vehicle applications is around 450~K. In order to maintain the required coercive field at this temperature neodymium (Nd) is partially replaced by heavy rare earth elements such as dysprosium (Dy). In the quest for reduction of critical elements including heavy rare earths elements methods for improving the coercive field of Nd$_2$Fe$_{14}$B magnets without Dy addition are sought. One of the strategies for high coercivity Dy-free magnets is the reduction of grain size. In order to produce aligned fine-grained, high coercivity magnets press-less sintering\cite{Sagawa2010} or hot-pressing of melt-spun ribbons with (Nd,Cu) infiltration\cite{SepehriAmin20136622} have been introduced.  

It is a well-known experimental fact that the coercive field of permanent magnets increases with decreasing grain size, however, there is little theoretical understanding for this.
This has to be attributed to the various factors which influence the reversal processes that change together with the grain size. The equation\cite{Kronmuller1988}
\begin{equation}
H_{\mathrm c}(T) = \alpha H_{\mathrm A}(T) - N_{\mathrm {eff}}M_{\mathrm s}(T) - H_{\mathrm{th}}(T)
\label{eq:alphaNeff}
\end{equation}
may be used to analyze magnetization reversal in permanent magnets. Here $H_{\mathrm A}$ is the anisotropy field and $M_{\mathrm s}$ is the saturation magnetization. The microstructural parameter $\alpha$ accounts for the reduction of the coercive field $H_{\mathrm c}$ owing to defects and misalignment. The effective demagnetizing factor $N_{\mathrm {eff}}$ accounts for the reduction of $H_{\mathrm c}$ by local magnetostatic interaction effects. It is, in general, not identical to the geometric demagnetizing factor and can be greater than 1. $H_{\mathrm{th}}$ includes the negative effects on coercivity owing to thermal fluctuations. The parameters $\alpha$ and $N_{\mathrm{eff}}$, which can be used to categorize permanent magnets, are extracted from the temperature dependent values of $H_{\mathrm c}(T)$, $H_{\mathrm A}(T)$, and $M_{\mathrm s}(T)$. This is done by plotting $H_{\mathrm c}(T)/M_{\mathrm s}(T)$ versus $H_{\mathrm A}(T)/M_{\mathrm s}(T)$ and fitting a straight line. When the thermal fluctuation field is not explicitly taken into account in the analysis, $N_{\mathrm{eff}}$, which is obtained from the intercept with the $H_{\mathrm c}/M_{\mathrm s}$-axis, also contains a term $H_{\mathrm{th}}/M_{\mathrm s}$.   

In this work, the grain size dependence of the coercive field is investigated.  In particular the focus is on Dy-free magnets. In hot-deformed magnets with (Nd,Cu) infiltration\cite{SepehriAmin20136622} it is possible to keep $\alpha$ constant and to control $N_{\mathrm {eff}}$. When selecting magnets for our experimental study of the grain size dependence of coercivity, we choose magnets with a similar $\alpha$ value. Therefore major differences in the coercive field between the investigated magnets have to be attributed to changes in the effective demagnetization factor due to local magnetostatic field effects.  By comparing experimental and computational results we show how the empirical relationship between coercive field and grain size can be understood using the theory of micromagnetism.  In order to compare the results with experimental data we compute a grain size dependent demagnetizing factor, $N^*$. Experiment and micromagnetic simulations give a demagnetizing factor that increases logarithmically with the grain size, $D$. Using a simple analytic model for $H_{\mathrm c}(D)$ we show that the key factor that causes the grain size dependence of the coercive field is the local demagnetizing field in the region where magnetization reversal starts.   

Several empirical relations between the grain size, $D$, and the coercive field, $H_{\mathrm c}$, have been proposed. Nucleation of reversed domains will be initiated near surface defects, which may be regions with low or zero magneto-crystalline anisotropy. Smaller grains have lower probability for surface defects. This argument was used to interpret the grain size dependence of the coercive field in sintered Nd$_2$Fe$_{14}$B magnets.\cite{Uestuener2006} With the same reasoning the size dependence of the coercive field was modelled by fitting the coercive field, $H_{\mathrm c}$, to the logarithm of the surface area, $S$. With $S$ being proportional to $D^2$, this gives the relationship\cite{Ramesh1988}
$H_{\mathrm c} = a - \tilde b \ln(D^2)$ 
or  
$H_{\mathrm c} = a - b \ln(D)$, using $b = 2 \tilde b$ where $a$ and $b$ are fitting parameters found using the experimental data. 

This model was applied by Li and co-workers\cite{Li2009} to fit the size dependence of the coercive field in sintered Nd$_2$Fe$_{14}$B magnets in the range from $D = 4.5~\mu$m to $D = 7.5~\mu$m. Alternatively, power laws of the form 
$H_{\mathrm c} = cD^{-d}$, where $c$ and $d$ are also fitting parameters found using the experimental data, were introduced to describe the grain size dependence of coercivity. Such a power law was used to describe the coercive field of sintered magnets with $D$ ranging from $3.5~\mu$m to $7.5~\mu$m.\cite{Uestuener2006} Weizhong and co-workers\cite{Weizhong1991} introduced the model
$H_{\mathrm c} = N_{\mathrm e}(\tilde \alpha H_{\mathrm A}-\tilde N_{\mathrm {eff}}M_{\mathrm s})$. This is a modified version of equation (\ref{eq:alphaNeff}) with the parameters $\alpha = N_{\mathrm e} \tilde \alpha$ and $N_{\mathrm {eff}} = N_{\mathrm e} \tilde N_{\mathrm {eff}}$, which are now correlated.
The factor $N_{\mathrm e}$ was found to be inversely proportional to the grain size. This again leads to a power law for $H_{\mathrm c}(D)$ with the prefactor $c = \tilde \alpha H_{\mathrm A}-\tilde N_{\mathrm {eff}}M_{\mathrm s}$. 

Gr\"onefeld and Kronm\"uller\cite{Gronefeld1989} computed the demagnetizing field near the edge of a Nd$_2$Fe$_{14}$B permanent magnet. From the local demagnetizing field they derived a local demagnetizing factor which they related to $N_{\mathrm {eff}}$ in equation (\ref{eq:alphaNeff}).  The transverse component of the demagnetizing field increases with decreasing distance towards the edge and becomes singular at the edge. Nevertheless, the magnetization near edges is found to be smooth as shown by analytic micromagnetic calculations.\cite{Thiaville1998} Similarly, numerical micromagnetic simulations show that the torque from the diverging demagnetizing field is balanced by the torque from the diverging exchange field near the edge.\cite{rave1998corners} At a point with a fixed distance from the edge the transverse component of the demagnetizing field increases logarithmically with the size of the cube.  

Micromagnetic computations of the grain size dependence of coercivity have been previously reported by various authors. Schabes and Betram\cite{Schabes1988} clearly show how the non-uniform demagnetizing field in a ferromagnetic cube causes the magnetization to rotate out of the uniaxial anisotropy direction near the edges and corners forming the flower state. Their study  was for particulate recording. The particles were in the size range from 5 nm to 55 nm. The moderate magnetocrystalline anisotropy of particulate media is much smaller than that of high performance permanent magnets. Schmidts and Kronm\"uller\cite{Schmidts1991} computed the coercive field of hard magnetic parallelepipeds using the intrinsic magnetic properties of Nd$_2$Fe$_{\mathrm 14}$B. Their simulations were two-dimensional. They assumed translational symmetry in a single direction perpendicular to the easy axis.  They included experimental data on the grain size dependence for melt-spun magnets and sintered magnets. However, with grain size several other parameters change so that a clear comparison of experiment with simulation was not possible.  For small particle sizes with $D < 100$~nm the numerically calculated coercive field decreases with a power law,  whereas for $D > 1000$~nm the numerically calculated coercive field decays logarithmically. Thielsch and co-workers\cite{Thielsch2013} showed that the local demagnetizing field of Nd$_2$Fe$_{14}$B grains, which have size and shape typically found in a hot-deformed magnet, causes a shape dependence of the coercive field. The total magnetic field, which is the sum of the external field and the demagnetizing field, initiates the nucleation of a reversed domain at the center of an edge. Near the edge where the nucleation starts the transverse component of the demagnetizing field means there is a finite angle between the total field and the easy axis even when the external field is applied at zero angle. The coercive field decreases with the angle of the total field according to the Stoner-Wohlfarth angular dependence.  The authors concludes that the local demagnetizing field at this very point in the magnet significantly influences the size dependence and shape dependence of coercivity.
Sepehri-Amin and co-workers \cite{sepehri-amin_micromagnetic_2014} used micromagnetics simulations of poly-crystalline model magnets to explain the grain size dependence of coercivity in anisotropic Nd-Fe-B sintered magnets. In particular, they showed that the magneto-static interaction fields of reversed surface grains increase with increasing grain size. 

In this work we focus on magnetization reversal of a single grain. 
We show that the logarithmic decay of the coercive field with grain size can be understood by the local self-demagnetizing field of a single grain. 
This field is non-uniform and initiates the nucleation of reversed domains near the edge of the grain. Thus the local magnetostatic field dominates over the macroscopic shape effect.  

\section{Methodology}

We use three-dimensional  finite element simulations\cite{exl2014} to compare micromagnetic theory with experiments. We include melt-spun magnets, (Nd,Cu) infiltrated magnets and sintered magnets in our study, covering a wide range of grain sizes (55 nm to 8300 nm). 
(Nd,Cu) infiltrated magnets have an intermediate grain size and the grains are almost perfectly isolated by a grain boundary phase. The composition, average grain size, coercive field and $\alpha$ and $N_{eff}$ parameters at room temperature of the Dy-free magnets used in our study are given in table~\ref{table:magnets}. 
The grain sizes were measured with Scanning Electron Microcopy (SEM) or Transmission Electron Microscopy (TEM) using the line intercept method.\cite{heyn1903short} The lines to estimate the average grain size were drawn perpendicular to the direction of the magnetic measurement. 
The measured mean grain diameter  is smaller than the actual grain size because lines do not necessarily intersect at the center of the grain, therefore we correct the grain size with a compensation formula\cite{fryxell1964creep,hensler1968relation}.

\emph{Sample~1} is a melt-spun ribbon with NdCu infiltration. The NdCu alloy with eutectic composition was infiltrated in the rapidly quenched ribbon at 580$^\circ$C for 60~min. The weight ratio of NdCu was 40\%. \emph{Sample~2} is an as melt-spun ribbon with high Nd, B concentration. The wheel speed was 20~m/s. \emph{Sample~3} is an isotropic sintered magnet produced from a rapidly quenched ribbon at 650$^\circ$C and a pressure of 400~MPa. The NdCu alloy with eutectic composition was infiltrated at 580$^\circ$C for 180~min.
\emph{Sample~4} was produced from hot pressing of melt-spun ribbons at 650$^\circ$C and a pressure of 400 MPa.  Die-upsetting was carried out with a strain rate of 1 s$^{-1}$ at 780$^\circ$C. Furthermore, the NdCu alloy with eutectic composition was infiltrated into the hot deformed magnet at 580$^\circ$C for 180~min. The weight ratio of NdCu was 20\%. \emph{Sample~5} was produced from hot pressing of melt-spun ribbons at 650$^\circ$C at a pressure of 400 MPa.  Die-upsetting was carried out with a slow strain rate of 0.01 s$^{-1}$ at 780$^\circ$C. Furthermore, the NdCu alloy with eutectic composition was infiltrated into the hot deformed magnet at 580$^\circ$C for 180~min.
The weight ratio of NdCu was 40\%. \emph{Sample~6} is a commercial Dy-free sintered magnet. 
Samples 1 to 3 are isotropic magnets and samples 4 to 6 are aligned magnets.

\begin{table}[t]
\caption{Magnets used for analyzing the grain size dependence of coercivity.} 
\label{table:magnets} % is used to refer this table in the text
\centering % used for centering table
\begin{tabular}{l l r c c c} 
\hline\hline %inserts double horizontal lines
  & composition (at \%)  & $D$(nm) & $\mu_0H_{c}$(T) & $\alpha$ & $N_{\mathrm {eff}}$\\ \hline
1 & %melt spun ribbon with infiltration& 
Nd$_{13.9}$Fe$_{75.5}$Co$_{4.5}$B$_{5.5}$Ga$_{0.6}$ & 
55 & 2.66 & 0.40 & 0.20 
\\
2 & %as melt-spun: Nd, B rich & 
Nd$_{15}$Fe$_{70}$B$_{14}$Ga$_1$ & 
60 & 2.83 & 0.42 & 0.25  
\\
3 & %sintered ribbon with infiltration & 
Nd$_{13.9}$Fe$_{75.5}$Co$_{4.5}$B$_{5.5}$Ga$_{0.6}$ & 
88 & 2.61 & 0.43 & 0.38
\\ 
4 & %hot deformed magnet with infiltration (small grains) &
Nd$_{13.9}$Fe$_{75.5}$Co$_{4.5}$B$_{5.5}$Ga$_{0.6}$ & 
346 & 2.41 & 0.48 & 0.76
\\
5 & %hot deformed magnet with infiltration (large grains) & 
Nd$_{13.9}$Fe$_{75.5}$Co$_{4.5}$B$_{5.5}$Ga$_{0.6}$ & 
789 & 1.95 & 0.45 & 0.91
\\
6 & %Dy free sintered magnet & 
Nd$_{14.6}$Fe$_{76.9}$Co$_{1.8}$B$_{6.1}$Al$_{0.5}$Cu$_{0.1}$ & 
8326 & 1.40 & 0.45 & 1.22 
\\ 
\hline 
\end{tabular}
\end{table}

Using an energy minimization method\cite{exl2014} we computed the coercive field as a function of particle size for three different shapes: The sphere, the dodecahedron and the cube, which are all perfect particles without any defects. The intrinsic material properties are kept uniform within each particle. We take the following intrinsic magnetic properties for Nd$_2$Fe$_{14}$B: Magnetocrystalline anisotropy constant $K_1 = 4.9$~MJ/m$^3$, spontaneous magnetization $\mu_0M_{\mathrm s} = 1.61$~T, and exchange constant $A = 8$~pJ/m. The Bloch wall width, $\delta_{\mathrm B}=\pi\sqrt{A/K_1}$, is 4~nm and the exchange length, $L_{\mathrm {ex}}=\sqrt{A/(\mu_0M_{\mathrm s})}$, is 1.97~nm. We use a geometrically scaled tetrahedral mesh which is refined towards the edges of the magnet. Following Rave and co-workers\cite{rave1998corners} the mesh size along the edges was set to $L_{\mathrm {ex}}/2$.

The micromagnetic results will be compared with a simple analytic model that applies the Stoner-Wohlfarth theory locally in the region where magnetization reversal is initiated.\cite{Thielsch2013} We evaluate the demagnetization field of a uniformly magnetized cube with edge length $D$ using the equation given by Akoun and Yonnet\cite{Akoun1984}. We give these equations in the appendix.

\section{Results and discussion}

\begin{figure}[t]
\includegraphics[width=0.85\columnwidth]{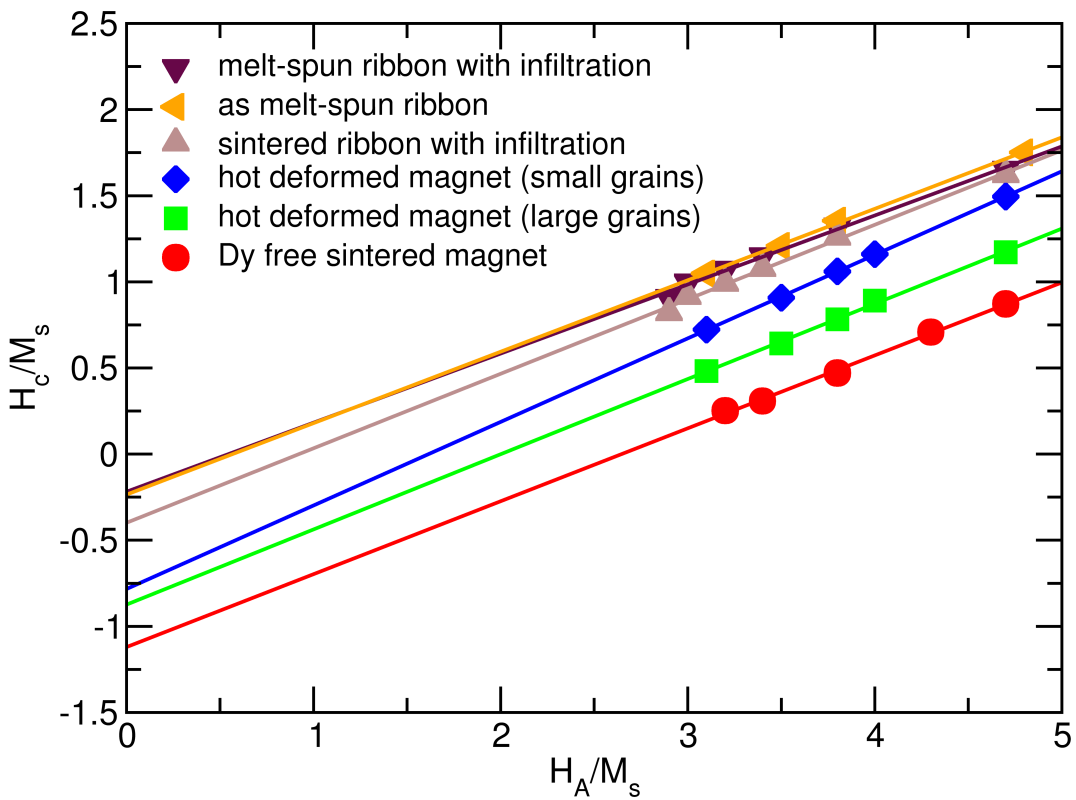}
\caption{Fit of the temperature dependent coercive field, in order to estimate the microstructural parameters for the magnets listed in table \ref{table:magnets} .}
\label{fig:KronmFit}
\end{figure}

Figure \ref{fig:KronmFit} shows a plot of the normalized coercive field $H_{\mathrm c}/M_{\mathrm s}$ versus $H_{\mathrm A}/M_{\mathrm s}$ evaluated at different temperatures. The data can be fitted to a straight line. According to equation (\ref{eq:alphaNeff}) the slope  gives the microstructural parameter $\alpha$ and the intercept with the $y$ axis gives the effective demagnetization factor $N_{\mathrm {eff}}$.  The temperature dependent values for $H_{\mathrm A}$ and $M_{\mathrm s}$ were taken from Gr\"ossinger and co-workers\cite{Groessinger1985} and Hock\cite{Hock1988}, respectively. The lines in figure \ref{fig:KronmFit} are almost parallel to each other which indicates that the $\alpha$ values of the magnets are similar. The sequence of lines from top to bottom starts with the melt-spun ribbons followed by the infiltrated hot-deformed magnets and the Dy free sintered magnet at the bottom. This reflects the increase of the $N_{\mathrm {eff}}$ with increasing grain size. The microstructural parameters calculated by fitting the experimental data are listed in table \ref{table:magnets}.  

\begin{figure}[t]
\includegraphics[width=0.85\columnwidth]{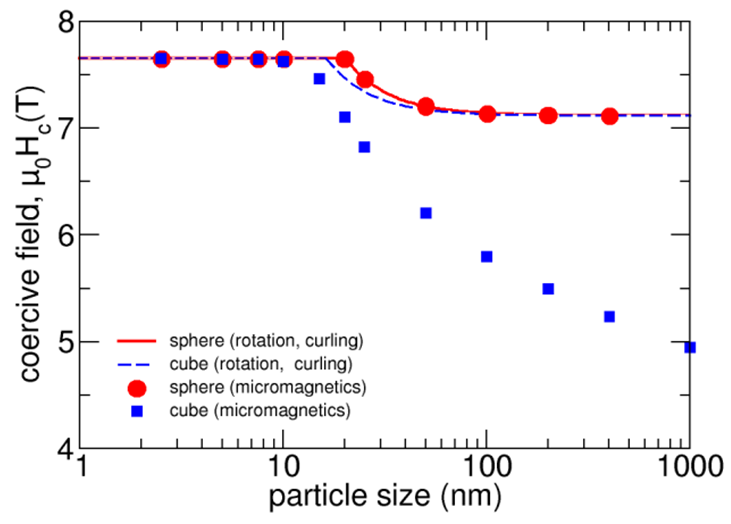}
\caption{Computed nucleation field of a Nd$_2$Fe$_{14}$B sphere (circles) and a Nd$_2$Fe$_{14}$B cube (squares) as function of size. Solid line: Theoretical nucleation field of a sphere\cite{Skomski2003}.  Dashed line: Theoretical nucleation field of a cube\cite{Aharoni1999}. The $x$-axis gives the diameter of the sphere or the edge length of the cube.}
\label{fig:SphereCube}
\end{figure}

Figure \ref{fig:SphereCube} compares the computed coercive field for a Nd$_2$Fe$_{14}$B sphere and a Nd$_2$Fe$_{14}$B cube. For the sphere, the micromagnetics results follow the theoretical nucleation field for uniform rotation at small diameters ($D < D_{\mathrm c}$) and for the curling mode at larger diameters ($D > D_{\mathrm c}$). The critical diameter $D_{\mathrm c}$ is 10.198~$L_{\mathrm {ex}}$. It is interesting to note that the coercive field of a sphere reaches a finite value $H_{\mathrm c} = 2K_1/(\mu_0 M_{\mathrm s})-(1/3)M_{\mathrm s}$ for $D\gg L_{\mathrm {ex}}$. 
In contrast, the computed coercive field of cubes decreases with increasing grain size for all $D > L_{\mathrm {ex}}$. The deviation from the theoretical prediction has to be attributed to the non-uniform demagnetizing field.

\begin{figure}[t]
\includegraphics[width=0.85\columnwidth]{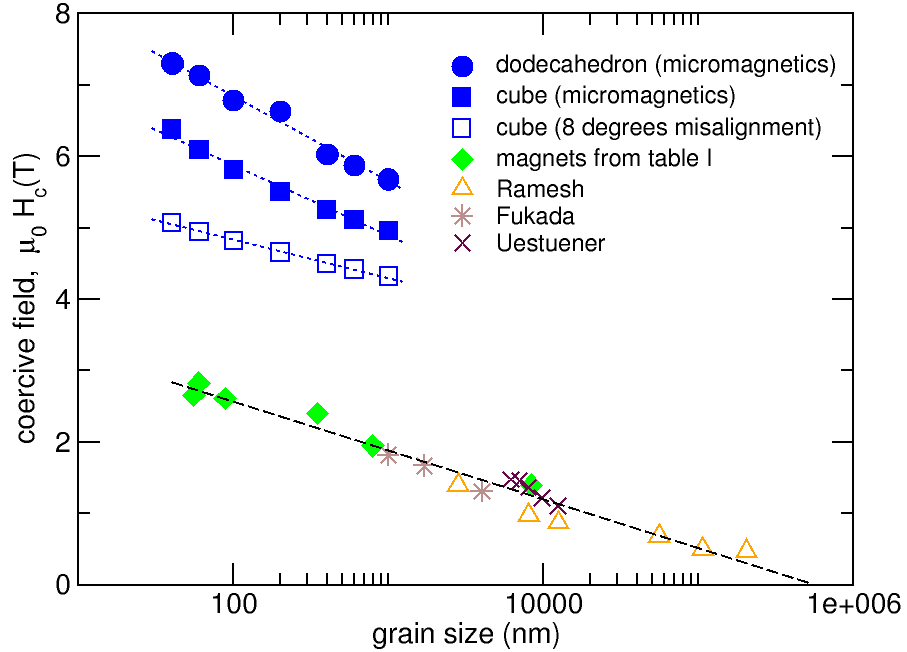}
\caption{Reversal field as function of grain size. \emph{Circles}: Numerical results for the dodecahedron. \emph{Full Squares}: Numerical results for the cube, perfectly aligned. \emph{Open Squares}: Numerical results for the cube, field angle of 8 degrees. \emph{Diamonds}: Experimental data for magnets of table \ref{table:magnets}. \emph{Triangles}: Experimental data by Ramesh and co-workers\cite{Ramesh1988}. \emph{Stars}: Fine grained sintered magnets by Fukada and co-workers\cite{Fukada2012}, \emph{Xs}: Sintered magnets by Uestuener and co-workers\cite{Uestuener2006}.}
\label{fig:hc}
\end{figure}

Figure \ref{fig:hc} shows the grain size dependence of the coercive field computed for the cube and the dodecahedron. In addition the plot contains the experimental data of the magnets in table \ref{table:magnets}  together with various other coercivity values for sintered magnets taken from literature\cite{Uestuener2006,Ramesh1988,Fukada2012}. 
The $\alpha$ values of the sintered magnets are expected to be comparable to those of hot-deformed NdCu infiltrated magnets. The dashed line in figure \ref{fig:hc} is a logarithmic fit to all experimental values. This result confirms the logarithmic decay of the coercive field with grain size, which is shown here for a wide range spanning several orders of magnitude. 
Similarly, the computed values for $H_{\mathrm c}(D$) decay logarithmically (dotted lines in figure \ref{fig:hc}). The switching field of the cube is lower than that of the dodecahedron. This reflects the difference in the demagnetizing field near the edges. At the same distance from the edge the demagnetizing field in the cube is higher than the demagnetizing field in the dodecahedron. 
The computed values for the coercive field are larger than the experimental values by more than a factor of two. This difference may be attributed to local defects in the magneto-crystalline anisotropy\cite{Kronmuller1988} or soft ferromagnetic grain boundary phases\cite{SepehriAmin20136622} not taken into account in the micromagnetic simulations. Further, thermal fluctuations, which are not taken into account in these  micromagnetic simulations, contribute to the offset between the calculated and experimental coercivities. Thermal fluctuations may help the system to overcome a finite energy barrier within the measurement time. The reduction of $H_{\mathrm c}$ by thermal fluctuations can be estimated by computing the size of the energy barrier to reversal as a function of the applied field in order to estimate the field required to reduce the energy barrier to a height of 25 $\mathrm{k_{B}}T$. This method was successfully applied to compute thermally induced vortex nucleations in permalloy elements \cite{Dittrich2005} and the temperature dependence of coercivity in magnetic recording media \cite{saharan2011angle}. In permanent magnetic grains the reduction in $H_{\mathrm c}$ owing to thermal jumps over energy barries is typically estimated at about 20 percent.\cite{bance2014influence}

Next we analyze the results using the following equation, which was suggested by Kronm\"uller and F\"ahnle\cite{Kronmuller2003},
\begin{equation}
H_{\mathrm c} = \alpha^*\frac{2K_1}{\mu_0M_{\mathrm s}}-n\ln\left(\frac{D}{\delta_{\mathrm B}}\right)M_{\mathrm s}.
\label{eq:Nsize}
\end{equation}

where $\alpha^{*}$ is an effective microstructural parameter. 
Kronm\"uller and F\"ahnle showed that the nucleation field of free NdFeB particles follows the logarithmic law (Equation \ref{eq:Nsize}).
They also reported that the effective demagnetization factor $N_{\mathrm{eff}}=n\ln(D/\delta_{\mathrm{B}})$ from micromagnetics simulations agrees well with experimental results. 
The finding by Kronm\"uller and F\"ahnle clearly shows that the grain size dependence in nucleation-controlled permanent magnets is a magnetostatic effect, which leads to a logarithmic decay of coercivity with grain size. The logarithmic law (Equation \ref{eq:Nsize}) has to be distinguished from the coercivity resulting from statistical pinning theory, which would lead to a coercivity proportional to $\sqrt{\ln D}$.\cite{Kronmuller2003} 
We believe that the magnets from which we obtained the experimental data for comparison in this paper are nucleation controlled. The $\alpha$ values (Table \ref{table:magnets}) are all greater than or equal to 0.4 which, according to Kronm\"uller and co-workers \cite{Kronmuller1988}, strongly indicates that pinning plays no role in the coercivity mechanism. Therefore we restrict our discussion to a single-grain defect-free magnet. 
Interestingly, this is sufficient to explain the experimentally-found logarithmic decay of coercivity. 
From the slope, $n$, of the curve $H_{\mathrm c}/M_{\mathrm s}$ versus $\ln(D/\delta_{\mathrm B})$ we can derive a size dependent demagnetizing factor
\begin{equation}
N^* = n\ln\left(\frac{D}{\delta_{\mathrm B}}\right).
\label{eq:Nstar}
\end{equation}

\begin{figure}[t]
\includegraphics[width=0.85\columnwidth]{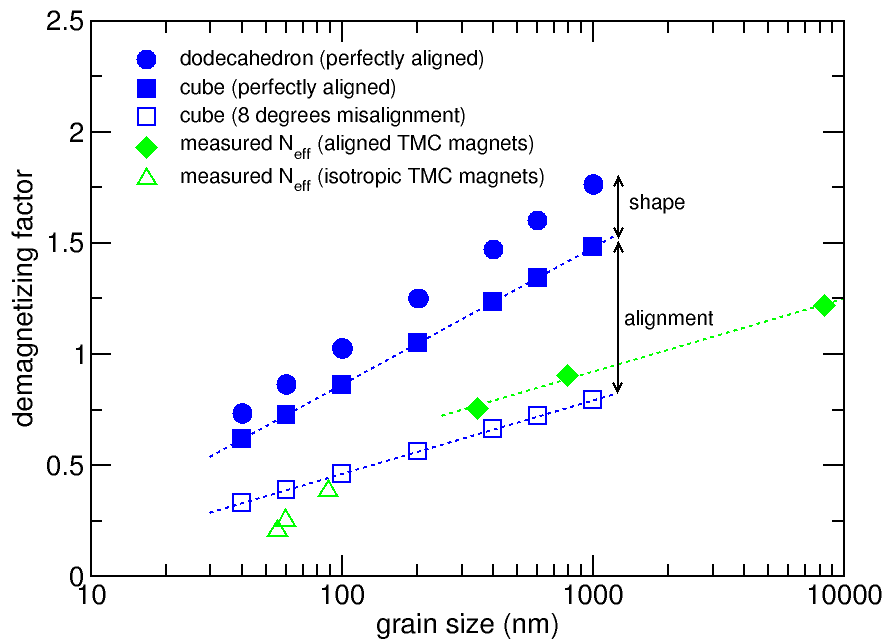}
\caption{Size dependent demagnetization factor, $N^*$, computed for a dodecahedron (\emph{circles}) and the cube (\emph{full squares}: perfectly aligned, \emph{open squares}: 8 degree field angle). Experimentally measured effective demagnetizing factor $N_{\mathrm {eff}}$ for the aligned magnets (\emph{diamonds}) and the isotropic magnets (\emph{triangles}) of table \ref{table:magnets}. The arrows indicate the change of slope caused by the differences in shape and alignment of the particles. }
\label{fig:demagfactor}
\end{figure}

The demagnetizing factor increases with the grain size. The factor $n$ in Equation (\ref{eq:Nstar}) is related to the slope of the $H_{\mathrm c}( \ln D)$  curve, which is given by $-nM_{\mathrm s}$. The numerical results (see Figure \ref{fig:demagfactor}) suggest that $n$ depends on the particle shape and the degree of alignment. The  values of  $n$ are 0.32, 0.27, and 0.14 for the aligned dodecahedron, the aligned cube, and the cube rotated by 8 degrees with respect to the external field, respectively. Using the room temperature values for the magnets in  table~\ref{table:magnets} we obtain  $n = 0.17$. Fitting all experimental values (our own data from table \ref{table:magnets} and the literature values\cite{Uestuener2006,Ramesh1988,Fukada2012}) gives $n = 0.18$. In addition, Figure \ref{fig:demagfactor} gives the effective demagnetizing factor of the aligned magnets (samples 3 to 6) of table \ref{table:magnets}. The plot clearly shows that  the  effective demagnetizing factor derived using Equation (\ref{eq:alphaNeff}) increases logarithmically with the grain size. Figure \ref{fig:demagfactor} also shows that the demagnetizing factor decreases if the grain is misaligned. Nevertheless the coercive field of the 
misaligned grain is smaller than that of the perfectly aligned sample. This indicates that the first term of Equation (\ref{eq:alphaNeff}) and the reduction of $\alpha$ with field angle is dominating. We conclude that the experimentally found increase of the coercive field with larger field angle has to be attributed to surface defects \cite{bance2014influence}. Such defects are not considered in this work where we restrict our numerical models on local demagnetizing effects of perfect particles.

In the following we develop a simple model that explains the logarithmic decay of the coercive field with grain size and the slope of $H_{\mathrm c}( \ln D)$. In particular we show that the demagnetizing field at the point where magnetization reversal starts is decisive for the grain size dependence of $H_{\mathrm c}$.

 For small cubes the drop of the coercive field as the field angle is changed from zero to eight degrees can be understood by the Stoner-Wohlfarth\cite{Stoner1948} theory. For a small sphere  which switches by uniform rotation the  switching field is
\begin{equation}
H_{\mathrm{sw}}  =  fH_{\mathrm A} \label{eq:stoner}
\end{equation}
\begin{equation}
f =  \left( \cos(\psi)^{2/3}+\sin(\psi)^{2/3}\right)^{-3/2} 
\label{eq:stonerfactor}
\end{equation}
where $\psi$ is the field angle. For $\psi=8$~degrees the reduction factor, $f$, is about 0.7. This partially explains the drop  of $H_{\rm c}$ between the perfectly aligned cube and the cube rotated by 8 degrees for small grain sizes. However the numerical simulation clearly show magnetization reversal by the nucleation and expansion of a reversed domain so that the Stoner-Wohlfarth theory is not directly applicable.

\begin{figure}[t]
\includegraphics[width=0.8\columnwidth]{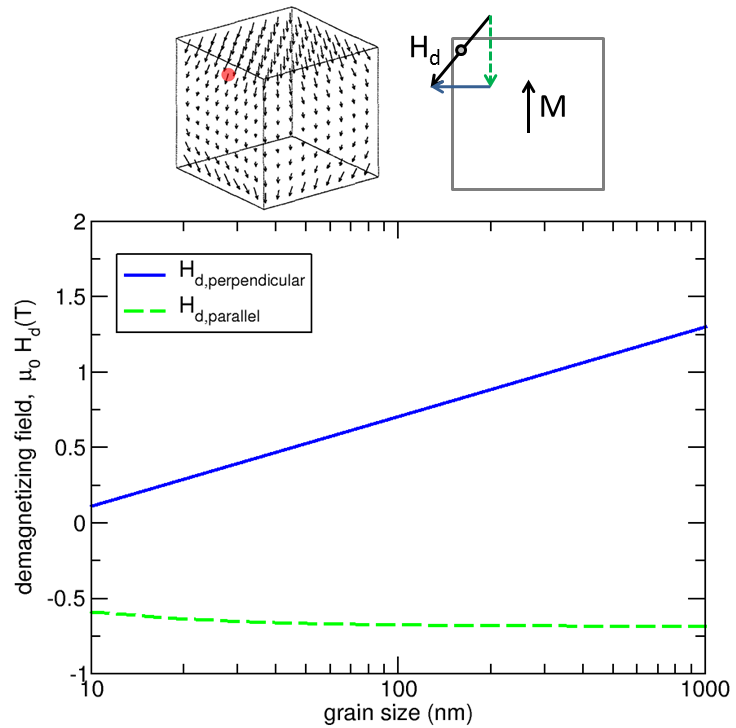}
\caption{Demagnetizing field of a uniformly magnetized cube evaluated at a distance of $d=1.2L_{\mathrm{ex}}$ from the edge. \emph{Solid line}: Component perpendicular to the easy axis. \emph{Dashed line}: Component parallel to the easy axis.}
\label{fig:hana}
\end{figure}

Following the arguments of Thielsch and co-workers\cite{Thielsch2013} we look at the angle of the total internal field in the region where magnetization reversal starts. They showed that Equation (\ref{eq:stoner}) can be applied locally, in order to estimate the coercive field, whereby $\psi\ $ is to
be replaced by the angle of the total internal field, $\psi_{\mathrm t}$, at the point where nucleation reversal starts. They found that in Nd$_2$Fe$_{14}$B particles with a rectangular prism shape magnetization reversal starts at the center of an edge. Next  we apply the Stoner-Wohlfarth theory locally within the reversal volume near the edge of the cube.

We use an analytic expression\cite{Akoun1984} for the demagnetizing field of a cube and evaluate the demagnetizing field at a distance $d = 1.2L_{\mathrm{ex}}$ from the center of an edge. Figure \ref{fig:hana}  shows the components of  $\mathbf{H}_{\mathrm d}$ parallel and perpendicular to the easy axis. With increasing size of the cube the perpendicular component of the demagnetizing field increases. This in turn leads to an increase of the angle $\psi_{\mathrm t}$. The external field is applied either parallel to the easy axis or at an angle of $\psi_{\mathrm{ext}} = 8$ degrees. The components of the external field, $\mathbf{H}_{\mathrm{ext}}$, are   
\begin{equation}
H_{\mathrm{ext},\parallel} = -H_{\mathrm{ext}}\cos(\psi_{\mathrm{ext}})
\end{equation}
\begin{equation}
H_{\mathrm{ext},\perp} = H_{\mathrm{ext}}\sin(\psi_{\mathrm{ext}}). 
\end{equation}
The components of the exchange field, $\mathbf{H}_{\mathrm{x}}$, in the reversal region are evaluated as   
\begin{equation}
H_{\mathrm{x},\parallel} = 0, \label{eq:hexparallel}
\end{equation}

\begin{equation}
H_{\mathrm{x},\perp} = \frac{1}{\mu_0 M_{\mathrm s}}\frac{A}{d^2}. \label{eq:hexperp}
\end{equation}

\begin{figure}
\includegraphics[width=0.85\columnwidth]{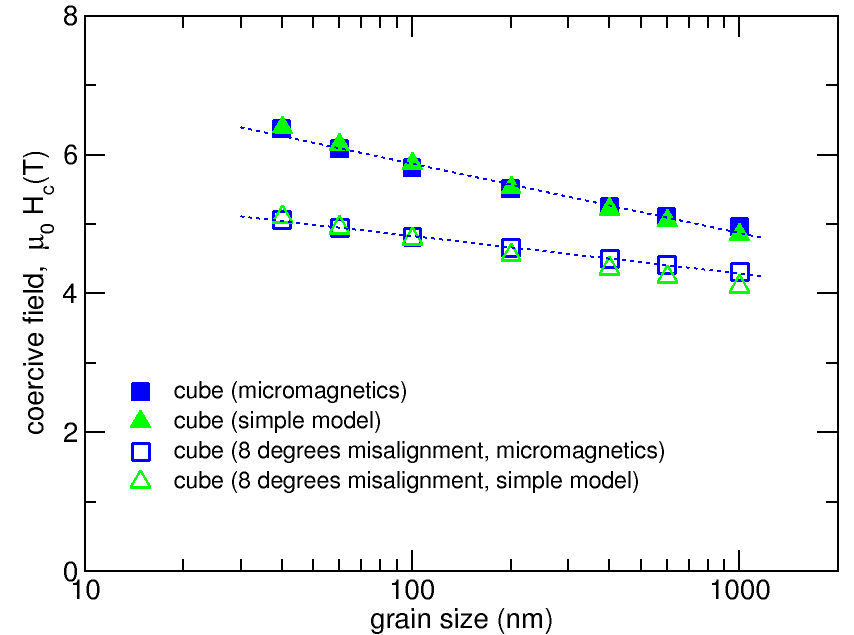}
\caption{Computed coercive field as a function of the grain size. Comparison of the full micromagnetic results (\emph{squares}) with a simple model (\emph{triangles}) for the coercive field. \emph{Full symbols}: Perfectly aligned cube. \emph{Open symbols}: Cube, field angle of 8 degrees.  }
\label{fig:simplemodel}
\end{figure}

\begin{figure}
\includegraphics[width=0.8\columnwidth]{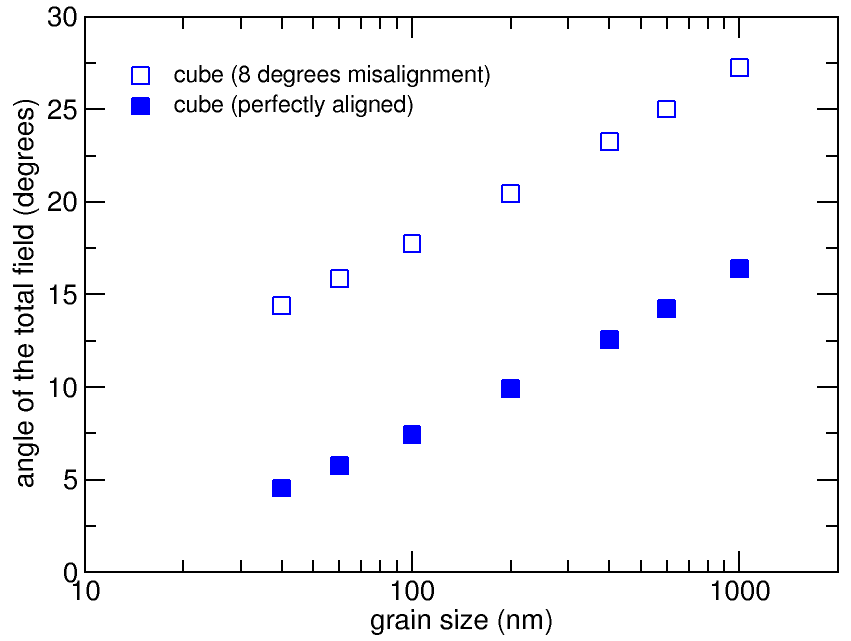}
\caption{Angle of the total internal field, $\psi_{\mathrm t}$, with respect to the easy axis at a distance of $d = 1.2L_{\mathrm{ex}}$ from the edge. \emph{Full squares}: Perfectly aligned cube. \emph{Open squares}: Cube,  8 degrees rotated.  }
\label{fig:angle}
\end{figure}

 The total internal field $\mathbf{H}_{\mathrm{t}}$ is the sum of the the demagnetizing field, the applied external field, and the exchange field:
\begin{equation}
\mathbf{H}_{\mathrm{t}} =  \mathbf{H}_{\mathrm{d}} + \mathbf{H}_{\mathrm{ext}} + \mathbf{H}_{\mathrm{x}}. 
\end{equation}
With the simple model we compute the coercive field as follows: We successively increase $H_{\mathrm{ext}}$. For each value of $H_{\mathrm{ext}}$ we evaluate  $\mathbf{H}_{\mathrm t}$ and compute $\psi_{\mathrm t}$. We denote the value of   $H_{\mathrm{ext}}$ when   
\begin{equation}
H_{\mathrm{ext}} \ge H_{\mathrm{sw}}(\psi_{\mathrm t})
\end{equation}
the approximate coercive field $\tilde H_{\mathrm c}$. Figure \ref{fig:simplemodel} shows that $\tilde H_{\mathrm c}$  coincides with the micromagnetically computed coercive field. In order to obtain a quantitative match between the simple model and the micromagnetic result, we have to include the exchange field, Equations (\ref{eq:hexparallel}) and (\ref{eq:hexperp}), in the simple model. This is different from the qualitative treatment by Thielsch and co-workers\cite{Thielsch2013} who only considered the external field and the demagnetizing field. In the simple model the only input that changes with the grain size is the perpendicular component of the demagnetizing field. With increasing grain size $H_{\mathrm d,\perp}$ increases. As a consequence the angle of the total internal field with respect to the easy axis increases. This is clearly seen in Figure \ref{fig:angle}, which shows $\psi_{\mathrm t}$ evaluated at
$H_{\mathrm{ext}} = \tilde H_{\mathrm c}$ for different grain sizes. For the perfectly aligned case the angles are between 4.5 degrees and 16 degrees. This range is shifted towards higher angles (14 degrees to 27 degrees) for the cube with 8 degrees misalignment. This difference explains the reduction of the slope of $H_{\mathrm c}( \ln D)$ in the case of a misalignment of 8 degrees. At small angles the factor $f$, see Equation (\ref{eq:stonerfactor}), changes rapidly with $\psi$. For larger angles $f(\psi)$ becomes flatter.\cite{Stoner1948}

The above model shows that the total field in the region where magnetization reversal starts determines the coercive field. In particular the angle of the total internal field with respect to the easy axis in the nucleation region, $\psi_{\mathrm t}$, is important. Due to their influence on the perpendicular component of the demagnetizing field a change in grain size or grain shape will change this angle. Clearly, the alignment of the grains will influence $\psi_{\mathrm t}$. Other microstructural effects that will have an influence at the local reversal conditions are  the nature of the grain boundary phase\cite{SepehriAmin20136622} and soft magnetic defects. Grain boundary phases may change the local exchange field and in turn alter $\psi_{\mathrm t}$. Defects in the local magneto-crystalline anisotropy will   change $H_{\mathrm A}$ in the nucleation region and thus modify $H_{\mathrm{sw}}$.
In a real magnet the interplay of various microstructural effects will determine  the slope of $H_{\mathrm c}( \ln D)$.

\section{Conclusion}

We showed that the numerically computed coercive field as a function of grain size for a single Nd-Fe-B grain can be fitted perfectly to a logarithmic law, Equation (\ref{eq:Nsize}). In the simulations we assumed the most simple model to explain this effect, a single isolated particle without defects. In our model the logarithmic decay of the coercive field can be attributed neither to the density of surface defects nor to stochastic domain wall pinning. Therefore we conclude that the logarithmic decay in polyhedral grains results from magnetostatic edge effects. Indeed, the equations for the demagnetizing field show a logarithmic singularity near the edge of a grain (see Appendix).

In summary, we confirmed the  logarithmic decay of $H_{\mathrm c}$ with grain size  for a simple micromagnetic model and a wide range of Nd$_2$Fe$_{14}$B magnets. The results suggest that the logarithmic decay of coercive field with increasing grain size results from the logarithmic increase of the demagnetizing field near the edges of a grain.  At this very location the torque exerted by the local field onto the magnetization initiates the formation of a reversed domain. With increasing particle size the torque that rotates the magnetization out of the anisotropy direction becomes larger
and domain formation happens at lower external fields. The slope of  $H_{\mathrm c}($ln$D$) depends on  the microstructural features including the grain shape, the degree of alignment and most likely the nature of the grain boundary phases. 

\section*{Acknowledgments}

This work is based on results obtained from the
future pioneering program ``Development of magnetic material technology for
high-efficiency motors'' commissioned by the New Energy and Industrial Technology
Development Organization (NEDO). We  acknowledge the financial support
from the Austrian Science Fund (F4112-N13).

\section*{Appendix}

The components of the demagnetizing field
of a uniformly magnetized particle of cuboidal
shape and dimensions $2a$, $2b$, and $2c$ which
is magnetized along the $z$ axis are:

\begin{equation}
H_{\mathrm d,\parallel}(x,y,z) = \frac{M_{\mathrm s}}{4\pi}\sum_{i=0}^{1} \sum_{j=0}^{1} \sum_{k=0}^{1} (-1)^{i+j+k} A(\xi_{i},\eta_{j},\zeta_{k})
\end{equation}

\begin{equation}
H_{\mathrm d,\perp}(x,y,z) = \frac{M_{\mathrm s}}{4\pi}\sum_{i=0}^{1} \sum_{j=0}^{1} \sum_{k=0}^{1} (-1)^{i+j+k+1} L_{\eta}(\xi_{i},\eta_{j},\zeta_{k})
\end{equation}

where
\begin{equation}
\xi_{i} = x - (-1)^{i}a 
\end{equation}

\begin{equation}
\eta_{j} = y - (-1)^{j}b 
\end{equation}

\begin{equation}
\zeta_{k} = z - (-1)^{k}c
\end{equation}

and

\begin{equation}
\rho(\xi,\eta, z) = (\xi^{2} + \eta^{2} + z^{2})^{1/2}
\end{equation}

\begin{equation}
A(\xi,\eta,\zeta) = \arctan \bigg( \frac{\eta \xi}{\zeta \rho} \bigg)
\end{equation}

\begin{equation}
L_{\eta}(\xi,\eta,\zeta) = \log(\eta + \rho)
\end{equation}

Here $H_{\mathrm d,\parallel}$ and $H_{\mathrm d,\perp}$ are the components of the
field parallel and perpendicular to the magnetization.
The origin of the coordinate system is at the center
of the cube. Fig. \ref{fig:hana} gives the field close to the edge
%%in function of the edge 
of a cube with edge length
$D = 2a = 2b = 2c$.

% Create the reference section using BibTeX (when submitting to APL you can't use a bibtex file!):
\bibliography{grainsize}

% APL doesn't accept BibTeX so once your paper is finished you need to copy the contents of your .bbl file here: and then recompiled. 
% % \begin{thebibliography}{23}
% % \end{thebibliography}

\end{document}